# Reconstruction of Cardiac Cine MRI using Motion-guided Deformable Alignment and Multi-resolution Fusion

Xiaoxiang Han[1], Yang Chen[2], Qiaohong Liu[3,*], Yiman Liu[4,5], Keyan Chen[1], Yuanjie Lin[1], Weikun Zhang[1]

[1]*School of Health Science and Engineering, University of Shanghai for Science and Technology, Shanghai 200093, P.R.China*

[2]*Algorithm Team, ToolSensing Technologies Co., Ltd, Chengdu, 610095, P.R.China*

[3]*School of Medical Instruments, Shanghai University of Medicine and Health Sciences, Shanghai 201318, P.R.China*

[4]*Department of Pediatric Cardiology, Shanghai Children's Medical Center, School of Medicine, Shanghai Jiao Tong University, Shanghai, 200127, P.R.China*

[5]*Shanghai Key Laboratory of Multidimensional Information Processing, School of Communication & Electronic Engineering, East China Normal University, Shanghai, 200241, P.R.China*

Correspondence: Qiaohong Liu, No. 279, Zhouzhu Highway, Pudong New District, Shanghai, P.R.China. Tel.: +86 137 6183 3680, Email: liuqh@sumhs.edu.cn

## Abstract

**Purpose:** Cardiac cine magnetic resonance imaging (MRI) is one of the important means to assess cardiac functions and vascular abnormalities. Mitigating artifacts arising during image reconstruction, and accelerating cardiac cine MRI acquisition to obtain high-quality images is important.

**Methods:** A novel end-to-end deep learning network is developed to improve cardiac cine MRI reconstruction. First, a U-Net is adopted to obtain the initial reconstructed images in k-space. Further to remove the motion artifacts, the Motion-Guided Deformable Alignment (MGDA) module with second-order bidirectional propagation is introduced to align the adjacent cine MRI frames by maximizing spatial-temporal information to alleviate motion artifacts. Finally, the Multi-Resolution Fusion (MRF) module is designed to correct the blur and artifacts generated from alignment operation and obtain the last high-quality reconstructed cardiac images.

**Results:** At an 8× acceleration rate, the numerical measurements on the ACDC dataset are SSIM of 78.40% ± 4.57%, PSNR of 30.46 ± 1.22 dB, and NMSE of 0.0468 ± 0.0075. On the ACMRI dataset, the results are SSIM of 87.65% ± 4.20%, PSNR of 30.04 ± 1.18 dB, and NMSE of 0.0473 ± 0.0072.

**Conclusion:** The proposed method exhibits high-quality results with richer details and fewer artifacts for cardiac cine MRI reconstruction on different accelerations.

# Keywords:

# INTRODUCTION

MRI is a widely employed clinical adjunctive diagnostic tool, offering advantages such as non-invasiveness, absence of ionizing radiation, and multiple parameter acquisition. Nonetheless, due to physiological and hardware constraints, the examination speed of MRI is slower than some other imaging modalities. As an integral component of MRI technology, cine MRI can provide the continuous assessment of organ anatomy and the physiological-pathological mechanisms in both the temporal and spatial domains. In contrast to conventional two-dimensional static MRI images, cine MRI with more crucial information in the temporal dimension is extensively applied in cardiac imaging, cine angiography, and cardiovascular disease diagnosis. Specifically for cardiac cine MRI reconstruction, the repeat acquirement of multiple heartbeat cycles spends more acquisition time which can induce patient discomfort and inevitably produce motion artifacts. Therefore, accelerated cardiac cine MRI acquisition is necessary for clinical practice.

Subsampling k-space data during MRI is the prevalent approach to reduce scanning time. However, due to the violation of Nyquist's theorem, images directly reconstructed from zero-filled under-sampled k-space data by performing Fourier inverse transform suffer from aliasing artifacts and lower SNR. In the past, numerous methods for reconstructing fully sampled MRI signals included partial Fourier transform [1], parallel imaging [2], compressed sensing (CS) [3], low-rank matrix completion [4], and manifold learning [5]. Nevertheless, traditional methods above often suffer hyper-parameter selection and slow reconstruction speed due to iterative optimization. In recent years, significant advancements have been achieved in the field of deep learning, driving progress across multiple domains[6][7][8][9][10][11][12]. Deep learning-based reconstruction methods have been developed to address these problems. In 2016, Wang et al. [13] pioneered the integration of deep learning into MRI reconstruction tasks. They employed convolutional neural networks to establish mappings between a large number of under-sampled images and fully sampled images. Then high-quality reconstructed images could be obtained by inputting only the under-sampled images during prediction. Subsequently, the success of this work led to the emergence of numerous deep learning-based static MRI reconstruction algorithms [14][15][16][17][18].

Although significant progress has been made in static MRI reconstruction, few recently proposed methods are oriented to dynamic

MRI reconstruction based on deep learning techniques. Batchelor et al. [19] pioneered the field of motion-compensated MRI reconstruction. Odille et al. [20] used compressed sensing-based methods for dynamic cardiac MRI reconstruction under free breathing conditions. Chandarana et al. [21] applied similar techniques to reconstruct dynamic liver MRI under free-breathing conditions. Aviles-Rivero et al. [22] improved fast dynamic MRI reconstruction using multi-task optimization. Similar works refer [23][24]. The traditional CS-MRI techniques have some limitations, such as parameter selection, optimization calculation, and low acceleration factor. In contrast, deep learning-based methods have become popular in cine MRI reconstruction due to their accuracy and efficiency. For instance, Schlemper et al. [25] designed a deep cascaded CNN-based network with a data sharing (DS) layer to capture temporal correlations between different frames for cine MRI reconstruction. Qin et al. [26] proposed CRNN, which uses convolutional recurrent neural networks to handle dynamic signals and leverages bidirectional recurrent hidden connections across time series to capture spatiotemporal correlations. Küstner et al. [27] performed 3D cardiac cine MRI reconstruction based on 4D spatiotemporal convolution, but it suffered from a high computational cost. Sarasaen et al. [28] fine-tuned their network using static high-resolution MRI as prior knowledge. Additionally, some studies [29][30] consider motion information as a crucial prior in cine MRI to improve the reconstruction quality. Eldeniz et al. [31] explored a deep learning-based reconstruction of dynamic liver MRI under free breathing conditions. Pan et al. [32] proposed a learning-based method for motion-compensated MR reconstruction to efficiently deal with non-rigid motion corruption in cardiac MR imaging. However, the time-consuming computational cost of motion-compensated reconstruction makes it unsuitable for real-time requirements. Kunz et al. [33] developed an end-to-end model specifically for free breathing cardiac cine MRI reconstruction based on deep learning. More works refer [34][35]. Commonly, the end-to-end methods show higher computational efficiency which is more suitable for real-time requirements. Spatial-temporal information is the key to accessing dynamic image understanding. Thus, aiming at taking full advantage of spatial-temporal information in consecutive MRI frames to effectively alleviate the motion artifacts and faithfully reconstruct the cardiac structure, a new end-to-end network that combines the motion-guided deformable alignment module and multi-resolution fusion module is proposed for cardiac cine MRI reconstruction. The main contributions of this paper are as follows. (1) A new end-to-end deep learning network intending to reconstruct high-quality cardiac images with minor errors at high acceleration factors is developed to improve cardiac cine MRI reconstruction. (2) A novel motion-guided deformable alignment method with second-order bidirectional propagation that effectively utilizes spatiotemporal information in cine Magnetic Resonance Imaging (MRI) is proposed to mitigate the impact of motion artifacts. (3) A multi-resolution fusion module combines Transformer and CNN is designed to further correct alignment errors or artifacts that may arise.

# MATERIALS AND METHODS

## Problem statement

Aiming to reconstruct the fully sampled MRI data from under-sampled Cine MRI data, the problem can be formulated as follows:

$$y = M \odot Fx + \varepsilon$$

where $x$ represents the fully sampled MRI data, F denotes the Fourier transform, $M$ is the subsampling mask, $\odot$ represents pointwise multiplication, and $\varepsilon$ is the noise introduced during data acquisition.

To address this ill-posed inverse problem (1), deep learning is introduced to reconstruct the undersampled MRI data. The optimization formulation for this process is as follows:

$$\hat{x} = \underset{x}{\text{argmin}} \frac{1}{2} \|y - M \odot Fx\|_2^2 + \lambda L(\theta)$$

where $\hat{x}$ represents the reconstructed MRI data, $L(\theta)$ is the prior regularization term, and $\lambda$ is the regularization coefficient. In an end-to-end deep learning framework, the optimization process of $L(\theta)$ can be expressed as follows:

$$L(\theta) = \underset{\theta}{\text{argmin}} \frac{1}{2} \|x - f_{\text{DL}}(x_{\text{zf}}|\theta)\|_2^2$$

where $f_{\text{DL}}$ represents deep learning forward propagation, $\theta$ denotes the learnable parameters, and $x_{\text{zf}}$ represents zero-filled MRI data after subsampling.

**Overall architecture of the proposed method**

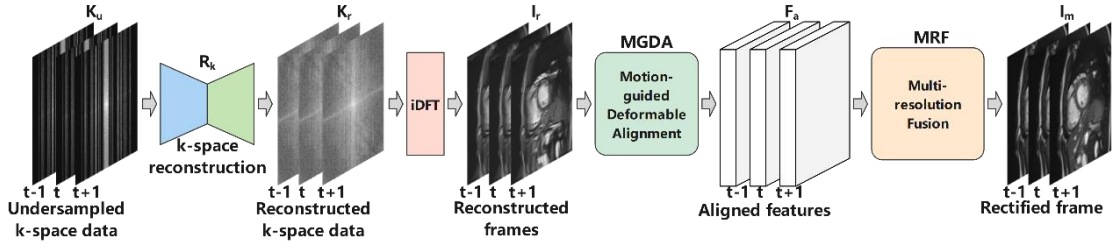

Fig.1. Overall architecture of the proposed method. The overall input of the network consists of multi-frame undersampled k-space data, while the output is the reconstructed fully sampled MR image. The proposed method's overall architecture comprises k-space reconstruction module, inverse discrete Fourier transform, motion-guided deformable alignment module, and multi-resolution fusion module.

As illustrated in Fig.1, a new end-to-end network associated with motion-guided deformable alignment (MGDA) module and multi-resolution fusion (MRF) module is proposed for free breathing cardiac cine MRI reconstruction. The proposed model comprises three main modules: the k-space signal reconstruction module, the motion-guided deformable alignment module, and the multi-resolution fusion module. Initially, the under-sampled k-space signals $K_u$ converted from consecutive cardiac cine MR images are fed into the k-space signal reconstruction module $R_u$ which adopts a traditional U-Net structure. The reconstructed k-space signals $K_r$ are converted into the image domain data $I_r$ by inverse Discrete Fourier Transform (iDFT). This process can be expressed as follows:

$$K_r = R_u(K_u)$$

$$I_r = \text{iDFT}(K_r)$$

Then, $I_r$ is processed by the MGDA module to obtain preliminary aligned feature maps $F_a$. Finally, $F_a$ is input into the MRF module for further processing to produce the final clear images. This process can be stated as:

$$F_a = \text{MGDA}(I_r)$$

$$I_m = \text{MRF}(F_a)$$

**Motion-guided deformable alignment module**

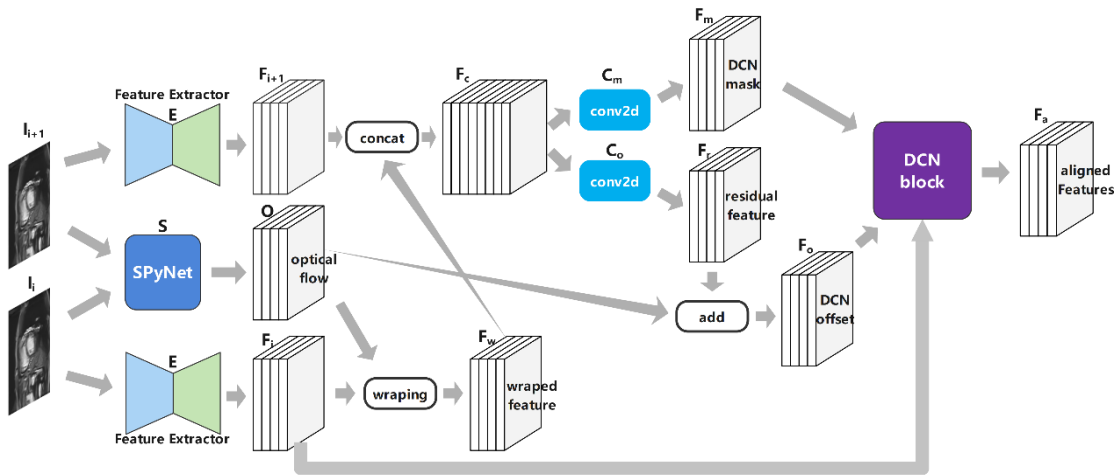

Fig.2. Motion-guided deformable alignment module. The MGDA consists primarily of the feature extractor, SPyNet for extracting optical flow information, and a deformable convolution network (DCN). This module generates all inputs for DCN and outputs aligned features.

During the reconstruction process, cardiac and respiratory motion, together with fast-flowing blood, may cause inter-slice motion artifacts. Thus, a motion-guided deformable alignment module is introduced to alleviate the effect of motion artifacts, as depicted in Fig.2. Deformable Convolutional Networks (DCN) [37][38] are used to align the adjacent cine MRI frames to eliminate the difference between these adjacent frames and achieve the goal of artifact removal. DCN refers to the addition of an offset at the sampling position in standard convolution operations, allowing the sampling grid to deform freely. Since these offsets should be learned from preceding layer features through an additional convolutional layer, deformation occurs in a local, dense, and adaptive manner conditioned on input features [39]. However, DCNs are challenging to train, and the training process is unstable, sometimes leading to offset overflow issues [40]. Therefore, motion information from adjacent frames can be utilized to guide the learning of offsets in DCN. A typical optical flow method named SPyNet [41] is employed to achieve the motion information estimation in the proposed MGDA module. SPyNet with strong performance, low computational cost, fast processing speed, and high compatibility can effectively capture motion information at different scales by processing the images in a spatial pyramid structure. Furthermore, SpyNet uses the pre-trained weights trained on a large-scale dataset to enhance the robustness of the proposed model. Additionally, the second-order grid propagation (SOGP) strategy [42] is adopted to aggregate more information from different spatial-temporal locations, aiming to improve the information aggregation ability in the network and the robustness of the network to occluded and fine regions, as shown in Fig.3.

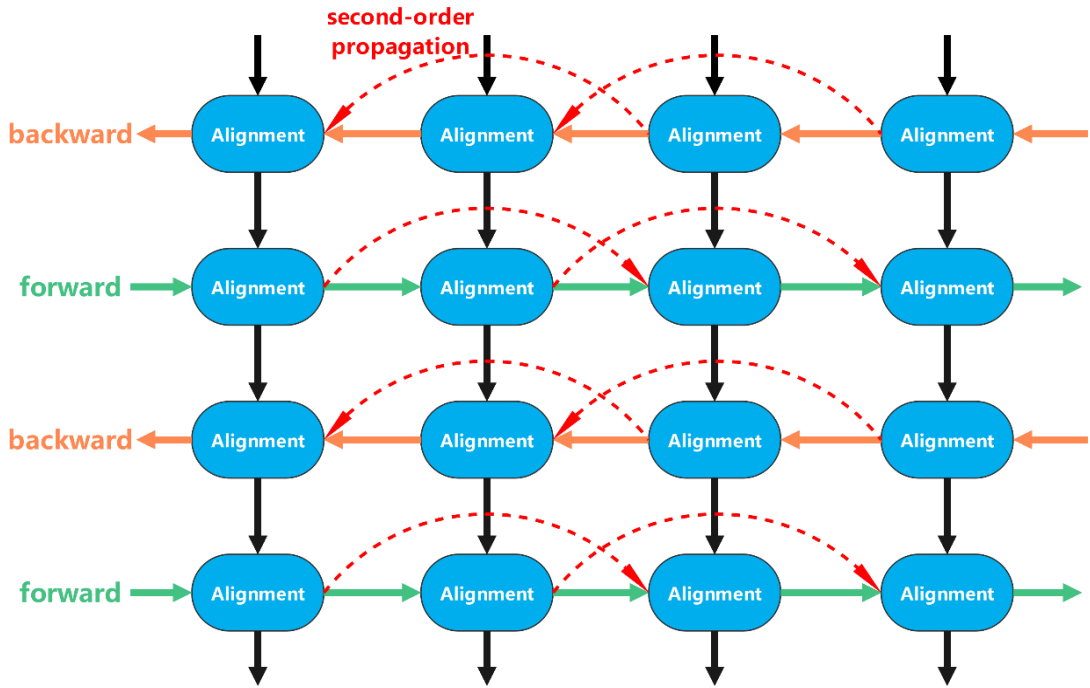

Fig.3. The second-order grid propagation strategy. The current state not only incorporates the hidden states of adjacent grid points but also those of points further away, facilitating effective removal of artifacts.

Firstly, the reconstructed images $I_i$ at frame $i$ and $I_{i+1}$ at frame $i+1$ are fed into the feature extractor $E$ respectively to obtain feature maps $F_i$ and $F_{i+1}$. Simultaneously, $I_i$ and $I_{i+1}$ are input into SPyNet for optical flow estimation, resulting in an optical flow field map $O$. Here, $E$ is a stacked residual block. The expressions for this process are as follows:

$$F_i = E(I_i)$$

$$F_{i+1} = E(I_{i+1})$$

$$O = S(I_i, I_{i+1})$$

where $S$ represents SPyNet. Subsequently, $F_i$ and $O$ perform a warping operation to obtain distorted features $F_w$. Then $F_w$ is concatenated with $F_{i+1}$ to obtain the feature map $F_c$. The expressions for this process are as follows:

$$F_w = \text{Wrap}(F_i, O)$$

$$F_c = \text{Cat}(F_{i+1}, F_w)$$

where $\text{Wrap}(\cdot, \cdot)$ denotes the warping operation, which mainly involves sampling the input features based on the deformed grid coordinates to obtain the deformed output. $\text{Cat}(\cdot, \cdot)$ represents the concatenation operation. Then, $F_c$ is calculated by two different convolutional layers $C_m$ and $C_o$ to obtain the feature maps $F_m$ and $F_r$, respectively. $F_r$ and $O$ are element-wise added to generate the feature map $F_o$. $F_m$ as the DCN mask, and $F_o$ as the DCN offset, with $F_i$ are fed into DCN block to yield the last aligned feature map $F_a$. The expressions for this process are as follows:

$$F_m = C_m(F_c)$$

$$F_r = C_o(F_c)$$

$$F_o = \text{Add}(F_r, O)$$

$$F_a = \text{DCN}(F_i, F_o, F_m)$$

where $\text{Add}(\cdot, \cdot)$ denotes element-wise addition, and $\text{DCN}(\cdot, \cdot, \cdot)$ represents the deformable convolutional network.

**Multi-resolution fusion module**

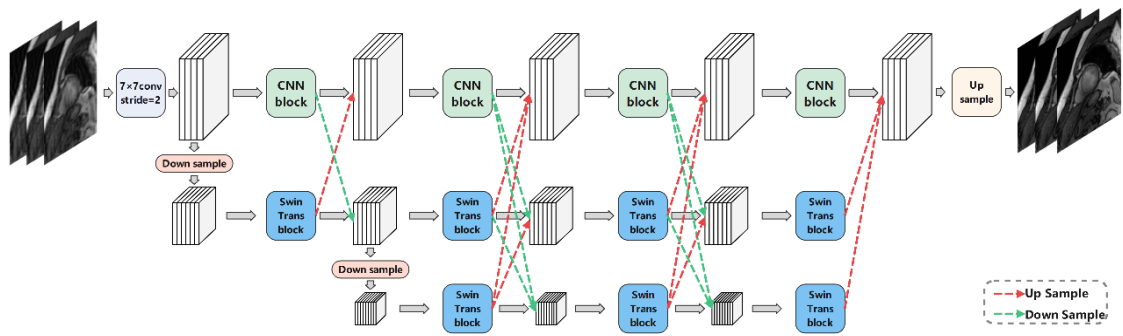

Fig.4. Multi-resolution fusion module. The module is divided into three branches with different resolutions. The high-resolution branch consists of CNN blocks, while the low-resolution branch consists of Transformer blocks.

Although the MGDA module can effectively alleviate the motion artifacts generated from cardiac, respiratory, and fast-flowing blood motion, the accuracy of alignment is difficult to ensure when there is significant motion between adjacent cine MRI frames, which would cause some unexpected artifacts like ghosting and blurring. Further, a new multi-resolution fusion (MRF) module that considers both inter-frame and intra-frame correlations is designed to remedy this problem. The overall architecture of MRF is illustrated in Fig.4. Inspired by HRNet's [43], the proposed MRF module consists of three parallel multi-resolution subnetworks that recurrently fuse representations generated by high to low subnetworks to obtain reliable high-resolution representations. Each low-resolution subnetwork is produced from the previous high-resolution subnetwork and high-resolution subnetwork features are progressively added to the corresponding low-resolution subnetwork features to form multiple stages with parallel connections between these multi-resolution subnetworks. Continuous exchange of information between three subnetworks through convolutional operations for multi-resolution fusion leads to richer high-resolution representations, which can effectively enhance the reconstructed details and spatial precision. Transformers [44], as a new network structure, can establish the long-range dependencies and capture temporal correlation in parallel, which benefits achieving more redundant information and enhancing the detail reconstruction. To obtain more advanced semantic information on the low-resolution subnetworks, the swin-Transformer [45] is used to capture additional global features, while simultaneously minimizing the increase of computational complexity. In the high-resolution subnetwork, the features from three different resolutions are fused to extract more effective high-level semantic features to produce the high-resolution output. Additionally, the MRF module employs 2×2 max-pooling with a stride of 2 for downsampling and 2× bilinear interpolation for upsampling.

## EXPERIMENTS AND RESULTS

### Datasets

The proposed method is validated on two publicly available cine MRI datasets, namely ACDC [46] and SACMRI [47]. The ACDC dataset, designed for automated cardiac diagnosis challenges, is created using real clinical exams conducted at the University Hospital in Dijon. It comprises 150 exams from different patients and is categorized into five classes, including normal cases, heart failure with infarction, dilated cardiomyopathy, hypertrophic cardiomyopathy, and right ventricular abnormality. The SACMRI dataset from The Hospital for Sick Children in Toronto consists of cardiac MR images from 33 subjects.

100 patients, 20 patients, and 30 patients of the ACDC dataset are utilized for training, validation, and testing, respectively. Each subject's data includes 12 to 35 frames and 6 to 21 cardiac slice data. 26 patients, 3 patients, and 4 patients of the SACMRI dataset are used for training, validation, and testing, respectively. Each patient's data comprises 20 frames and 8 to 15 cardiac slice data. Given the availability of only magnitude images in two datasets, the method described in reference [48] is selected to synthetically generate phase maps (smoothly varying 2D sinusoid waves) on-the-fly. This approach enhances the authenticity of the simulations by eliminating the conjugate symmetry within the k-space. Variable-density 1D under-sampling masks are employed with two different acceleration factors (4× and 8×) to obtain the corresponding zero-filling cardiac dataset.

**Implementation details**

This study is implemented using Python 3.8 and PyTorch 1.12.1 on a GPU server equipped with 1 Intel Core i9-10900X CPU, 32GB RAM, and 2 Nvidia RTX3080 (10GB) GPUs. Notably, an efficient and user-friendly PyTorch-Lightning 1.7.5 framework is leveraged to simplify the coding. The batch size is set based on data dimensions to maximize memory utilization. AdamW [49] is used as an optimizer with an initial learning rate of 1e-3 and ReduceLROnPlateau is adopted as the dynamic learning rate adjustment strategy. The model is trained for 100 epochs and the data loading program employs 16 threads. Automatic mixed precision is utilized during training and the mean squared error (MSE) loss is employed as the loss function. The performance of cine MRI reconstruction is evaluated by the Structural Similarity Index (SSIM) [50], Peak Signal-to-Noise Ratio (PSNR), and Normalized Mean Squared Error (NMSE).

**Experiments with other methods**

The proposed method is compared with some popular Cine MRI reconstruction techniques, including kt FOCUSS [51], DC-CNN [20], CRNN [26], DRN [30], MODRN [30], and RecurrentVarNet [52]. Among these compared methods, kt FOCUSS is a compressed sensing method, while the remaining methods are based on end-to-end deep learning techniques. MODRN is an improved version of integrating motion compensation into DRN.

The quantitative comparison of all methods on the ACDC dataset is presented in Table 1. The first row displays results for under-sampled images reconstructed directly from zero-filled k-space. The proposed method outperforms other methods at both 4× and 8× acceleration rates. The proposed method achieves excellent results closest to the ground truth, with the highest PSNR and SSIM and lowest NMSE values among all the methods. Specifically, at a 4× acceleration rate, the proposed method improves 1.36% in SSIM and 0.78dB in PSNR compared with the suboptimal model MODRN. At an 8× acceleration rate, the proposed method achieves a 1.71% improvement in SSIM and a 0.92dB improvement in PSNR compared with the suboptimal model MODRN. These quantitative measurements demonstrate that the proposed method has a promising performance for different acceleration rates.

Table 1 Quantitative comparison results on the ACDC dataset. (The best results are marked in bold.)

| Method | 4× | | | 8× | | |
|---|---|---|---|---|---|---|
| | PSNR(dB)↑ | SSIM(%)↑ | NMSE↓ | PSNR(dB)↑ | SSIM(%)↑ | NMSE↓ |
| zero-filled | 26.39 | 63.88 | 0.0868 | 24.70 | 55.31 | 0.1532 |
| kt FOCUSS | 29.14±1.26 | 74.26±4.25 | 0.0651±0.0075 | 26.02±1.27 | 61.84±5.18 | 0.1273±0.0088 |
| DC-CNN | 31.79±1.17 | 81.90±4.16 | 0.0389±0.0031 | 27.62±1.15 | 67.21±4.92 | 0.1023±0.0074 |
| DRN | 33.29±1.36 | 86.03±4.34 | 0.0298±0.0071 | 28.22±1.34 | 71.31±5.63 | 0.0884±0.0091 |
| CRNN | 33.95±1.38 | 87.24±5.14 | 0.0281±0.0062 | 28.85±1.14 | 72.38±5.54 | 0.0819±0.0086 |
| MODRN | 34.28±1.29 | 88.04±4.36 | 0.0258±0.0076 | 29.54±1.26 | 76.69±5.18 | 0.0626±0.0062 |
| RecurrentVarNet | 34.16±1.32 | 87.96±4.43 | 0.0269±0.0068 | 29.45±1.18 | 76.12±5.33 | 0.0687±0.0082 |
| MDAMF(ours) | **35.06±1.19** | **89.40±4.13** | **0.0238±0.0048** | **30.46±1.22** | **78.40±4.57** | **0.0468±0.0075** |

The comparisons of visualization results of the reconstruction under $4\times$ and $8\times$ acceleration conditions are shown in Fig.5 and Fig.6 respectively, with the reconstructed images, and their corresponding error maps using different methods on the ACDC dataset. As can be seen, the reconstructed results of the proposed method have best visual effects and least absolute errors relative to the ground truth compared with the other methods. The reconstructed images at the cardiac region of the proposed method have little artifacts and great details which benefit from the full utilization of spatial-temporal information in dynamic images.

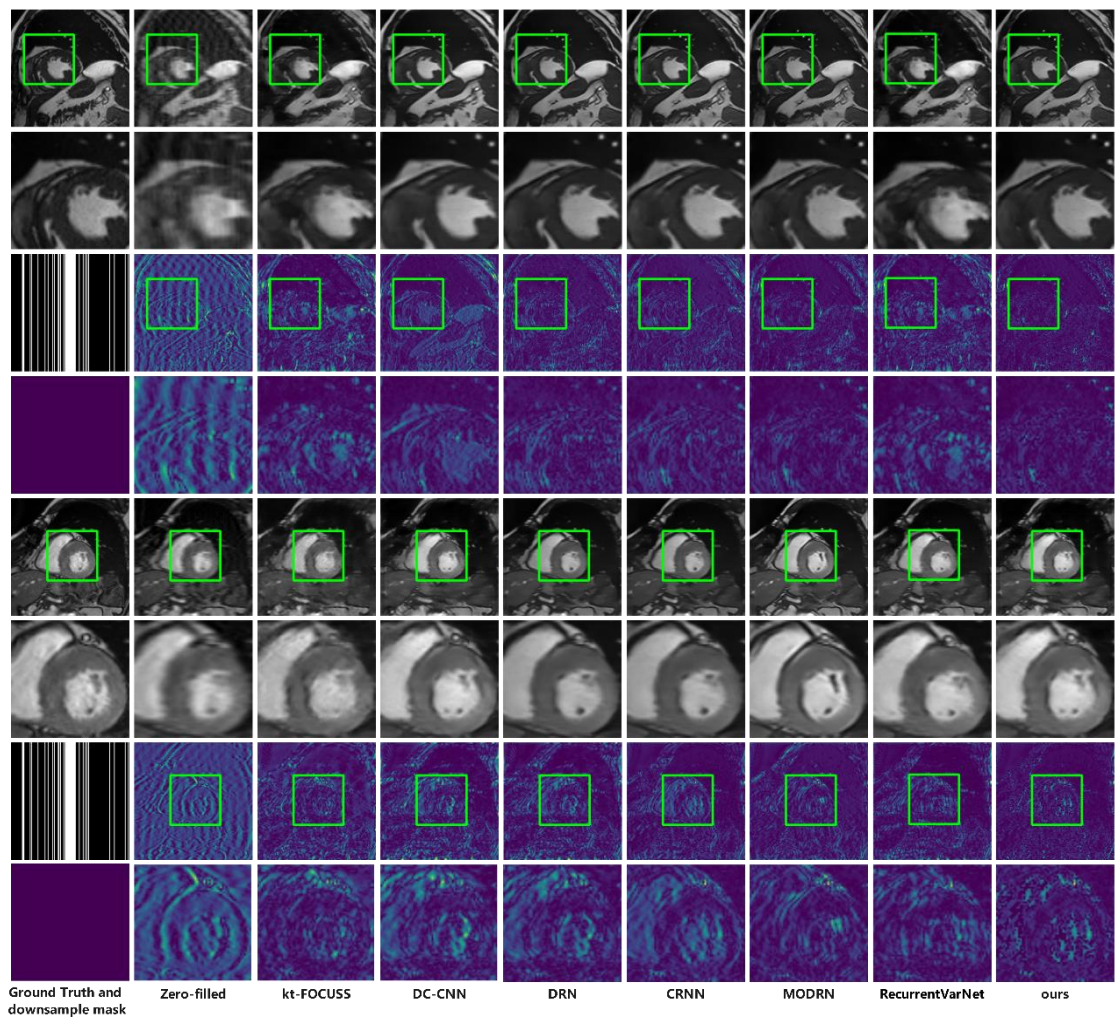

Fig.5. Qualitative results of different methods on the ACDC dataset under 4× acceleration.

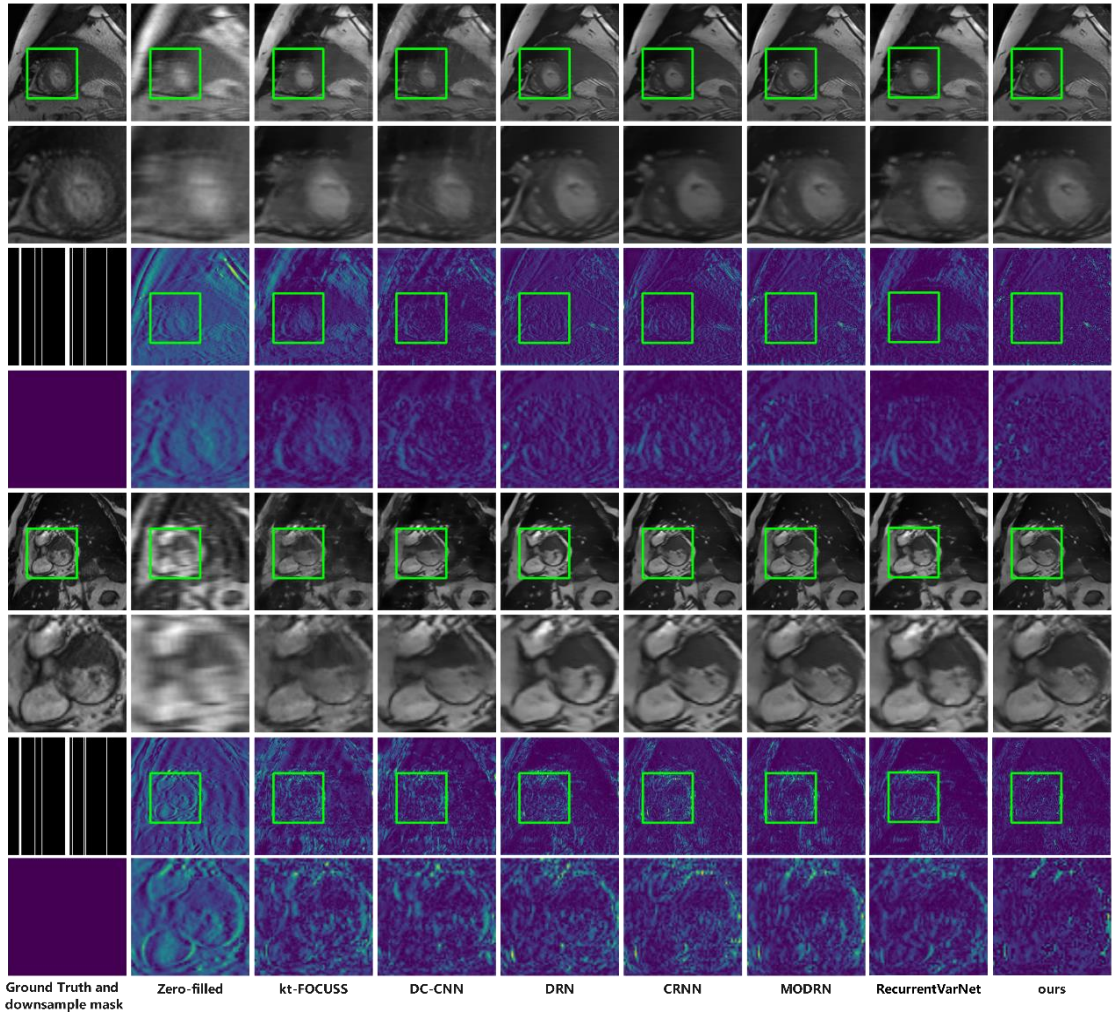

Ground Truth and downsample mask    Zero-filled    kt-FOCUSS    DC-CNN    DRN    CRNN    MODRN    RecurrentVarNet    ours

Fig.6. Qualitative results of different methods on the ACDC dataset under 8× acceleration.

    Furthermore, to verify the generalization of the proposed method, more comparison experiments are conducted on the SACMRI dataset under 4× and 8× acceleration conditions. The quantitative results of all methods under different acceleration factors are reported in Table 2. The proposed approach also outperforms other methods in terms of PSNR, SSIM, and NMSE values on the SACMRI dataset. Specifically, at a 4× acceleration rate, the proposed method improves 0.8% in SSIM and 0.94dB in PSNR compared with the suboptimal model MODRN. At an 8× acceleration rate, the proposed method improves 0.76% in SSIM and 0.61dB in PSNR relative to the suboptimal model MODRN. The

reconstructed images of different methods under 4× and 8× acceleration conditions on the SACMRI dataset are depicted in Fig.7 and Fig.8, respectively. The proposed method surpasses other methods with excellent quantitative and qualitative results, which further verify the generalization and robustness.

Table 2 Quantitative comparison results on the SACMRI dataset. (The best results are marked in bold.)

| Method | 4× | | | 8× | | |
|---|---|---|---|---|---|---|
| | PSNR(dB)↑ | SSIM(%)↑ | NMSE↓ | PSNR(dB)↑ | SSIM(%)↑ | NMSE↓ |
| zero-filled | 26.53 | 74.41 | 0.0871 | 25.15 | 56.94 | 0.1431 |
| kt FOCUSS | 29.48±1.79 | 86.26±4.58 | 0.0482±0.0075 | 26.79±1.54 | 78.56±5.12 | 0.0906±0.0085 |
| DC-CNN | 31.63±1.81 | 89.70±4.42 | 0.0287±0.0039 | 28.11±1.26 | 83.28±4.26 | 0.0798±0.0093 |
| DRN | 32.12±1.49 | 90.24±4.95 | 0.0272±0.0047 | 28.43±1.30 | 84.21±4.42 | 0.0664±0.0056 |
| CRNN | 32.42±1.62 | 91.08±5.13 | 0.0260±0.0051 | 28.98±1.44 | 85.11±4.83 | 0.0536±0.0042 |
| MODRN | 33.02±1.77 | 92.14±4.67 | 0.0248±0.0063 | 29.43±1.37 | 86.89±4.31 | 0.0496±0.0068 |
| RecurrentVarNet | 32.87±1.65 | 92.02±4.86 | 0.0257±0.0074 | 29.35±1.23 | 86.14±4.58 | 0.0510±0.0079 |
| MDAMF(ours) | **33.96±1.54** | **92.94±4.52** | **0.0235±0.0043** | **30.04±1.18** | **87.65±4.20** | **0.0473±0.0072** |

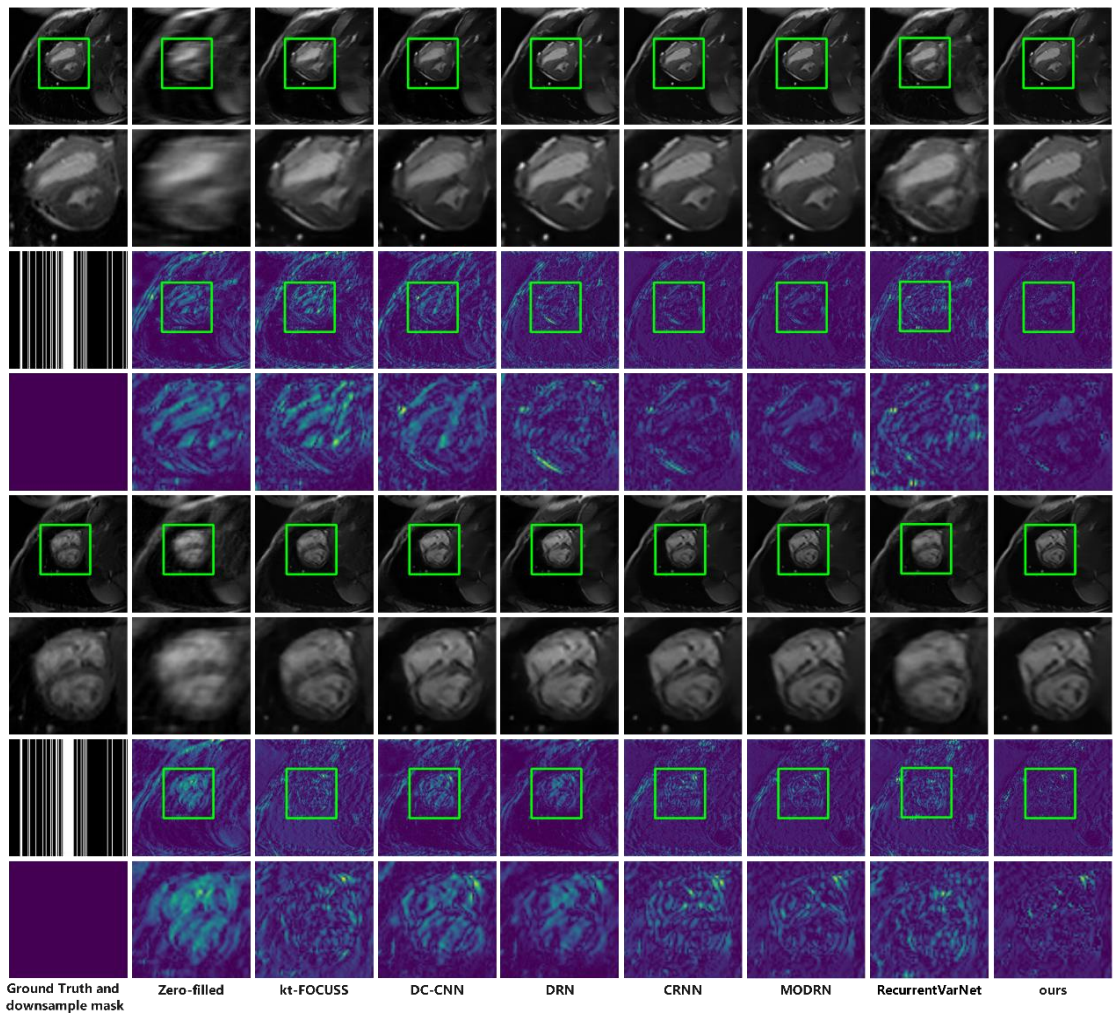

Fig.7. Qualitative results of different methods on the SACMRI dataset under 4× acceleration.

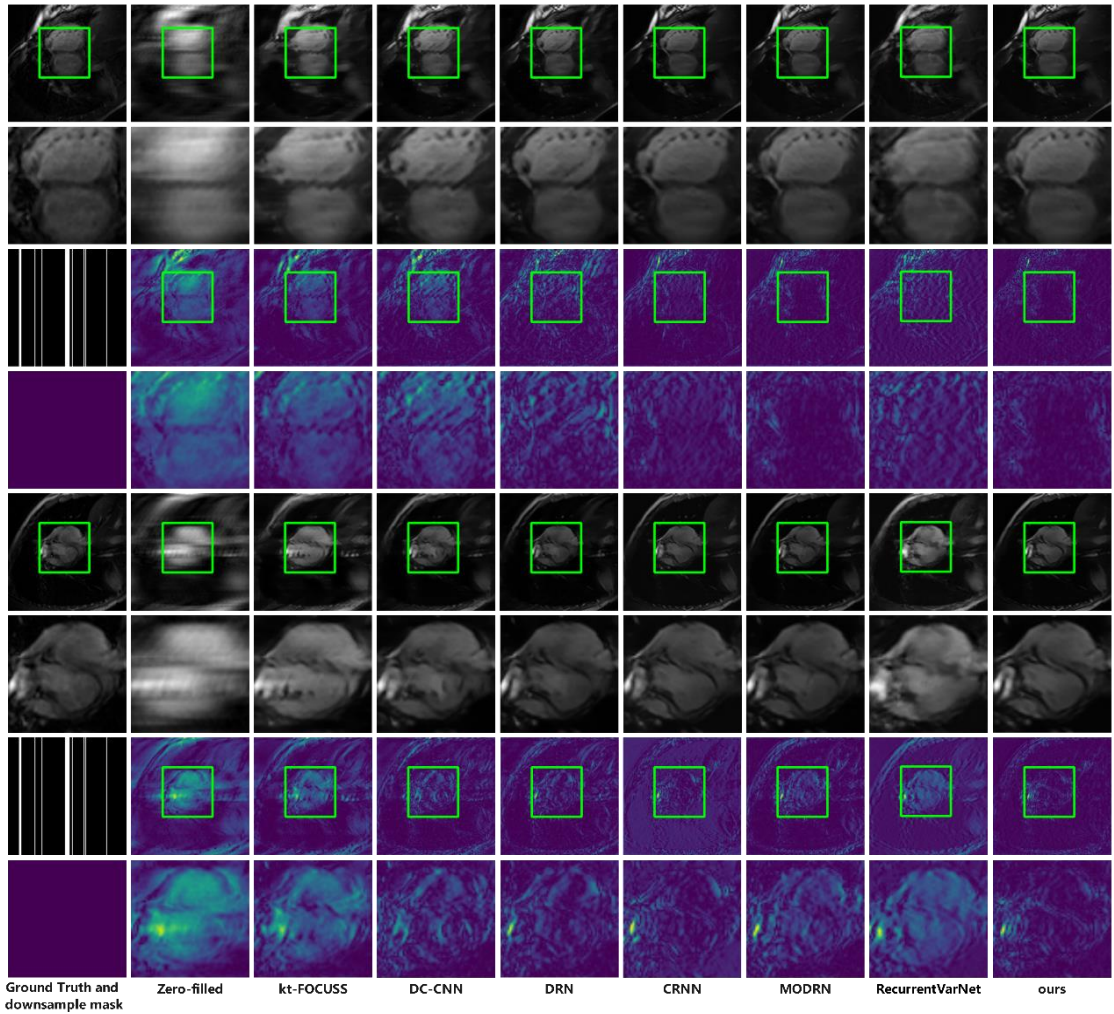

Fig.8. Qualitative results of different methods on the SACMRI dataset under 8× acceleration.

**Ablation study**

To demonstrate the feasibility and effectiveness of each component of the proposed network, three ablation studies are conducted based on the ACDC dataset. The first ablation study is to investigate the effectiveness of the MGDA and MRF modules whose compared results are listed in Table 3. Three metric values achieve the best performance by using both MGDA and MRF modules. Furthermore, the analyses of t-test between the proposed method and the model only using MGDA module, and the proposed method and the model only using MRF module are conducted to explain of differences. The t-test results indicate that the removal of these modules significantly differs from each other (p<0.001). MGDA compensates for motion artifacts in the current frame by capturing features from reference frames around, leading to a modest performance boost. Moreover, the potent feature fusion and artifact correction abilities of MRF contribute substantially to augmenting the model's reconstruction performance.

Table 3 Ablation study for verifying the effects of the MGDA and MRF module. (The best results are marked in bold.)

| MGDA | MRF | 4× | | | 8× | | |
|------|-----|------|------|------|------|------|------|
| | | PSNR(dB)↑ | SSIM(%)↑ | NMSE↓ | PSNR(dB)↑ | SSIM(%)↑ | NMSE↓ |
| ✓ | × | 34.12 | 87.98 | 0.0256 | 28.92 | 74.04 | 0.0765 |
| × | ✓ | 34.45 | 88.27 | 0.0248 | 29.54 | 75.93 | 0.0687 |
| ✓ | ✓ | **35.06** | **89.40** | **0.0238** | **30.46** | **78.40** | **0.0468** |

In the design MGDA module, second-order grid propagation (SOGP) strategy is adopted to aggregate more spatial-temporal information and improve the reconstruction performance. The second ablation study verifies the boosting ability between SOGP and first-order grid propagation (FOGP). As depicted in Table 4, SOGP facilitates the current frame in obtaining information from more different directional frames and exhibits greater competitiveness than FOGP. Additionally, the calculated t-test result can also demonstrate there is a significant difference between both cases (p<0.001).

Table 4 Ablation study for verifying the effects of the FOGP and SOGP of MGDA. (The best results are marked in bold.)

| FOGP | SOGP | 4× | 8× |
|------|------|-----|-----|

| | | PSNR(dB)↑ | SSIM(%)↑ | NMSE↓ | PSNR(dB)↑ | SSIM(%)↑ | NMSE↓ |
|---|---|---|---|---|---|---|---|
| ✓ | × | 34.86 | 88.48 | 0.0247 | 29.10 | 76.54 | 0.0598 |
| × | ✓ | **35.06** | **89.40** | **0.0238** | **30.46** | **78.40** | **0.0468** |

The last ablation study is about the architecture of the proposed MRF module, which consists of both CNN and Transformer. From Table 5, the experimental outcomes demonstrate that the performance is unsatisfactory when employing either CNN or Transformer in isolation, while the hybrid MRF architecture exhibits excellent performance. The t-test results between mutual two cases indicate that there are significant differences ($p<0.001$).

Table 5 Ablation study of different MRF architectures. (The best results are marked in bold.)

| CNN | Trans | 4× | | | 8× | | |
|---|---|---|---|---|---|---|---|
| | | PSNR(dB)↑ | SSIM(%)↑ | NMSE↓ | PSNR(dB)↑ | SSIM(%)↑ | NMSE↓ |
| ✓ | × | 34.57 | 88.90 | 0.0263 | 29.15 | 76.18 | 0.0608 |
| × | ✓ | 34.48 | 88.62 | 0.0258 | 29.29 | 76.40 | 0.0512 |
| ✓ | ✓ | **35.06** | **89.40** | **0.0238** | **30.46** | **78.40** | **0.0468** |

## DISCUSSION

All experiments in terms of qualitative or quantitative results consistently demonstrate the superiority, robustness, and generalization of the proposed method. To reconstruct the clear images with richer details and fewer artifacts from cardiac cine MRI, the proposed method designs MGAD and MRF modules based on the preliminary reconstructed results obtained using U-Net in k-space which can take full advantage of spatial-temporal characteristics of dynamic sequences. Meanwhile, the MGDA module uses DCN to align the adjacent cine MRI frames to eliminate the difference between these adjacent frames and remove the motion artifacts. In MGDA module, the second-order bidirectional propagation of SOGP strategy avoids the decay of alignment features from long-range alignments and acquits additional spatial positional features to refine detailed features. Finally, the MRF module effectively eliminates alignment errors, further removes the motion artifacts, and obtains the last reconstructed high-quality cardiac image.

Although the proposed method provides excellent results in dynamic cardiac cine reconstruction, it still has some limitations. First, it is essential to acknowledge potential limitations or biases in these datasets. One possible limitation could be the variability in image quality and characteristics among different subjects in the datasets, which may impact the generalizability of the results to a broader population. Additionally, the datasets may not fully represent all possible scenarios encountered in clinical practice, leading to potential biases in the evaluation of the proposed method's performance. Understanding these dataset characteristics is crucial for interpreting the results accurately and assessing the practical applicability of the proposed reconstruction method in real-world clinical settings. Second, the under-sampled data are synthetic. Therefore, more real-world data from clinical will be collected to verify the proposed method in the future. Third, the proposed method demands high computational resources. In future work, the complexity of the proposed method will be optimized to reduce computational costs. One promising direction is the exploration of alternative deep learning architectures to further enhance the performance and robustness of reconstruction methods. By investigating different architectures, researchers can potentially discover more efficient and effective ways to reconstruct cardiac images, leading to improved image quality and artifact reduction. Additionally, incorporating additional types of information into the reconstruction process could offer new opportunities for advancing the field. One potential area for future research is the integration of multi-modal information into the reconstruction process. By combining data from different imaging modalities, such as CT scans, researchers can potentially improve the accuracy and reliability of cardiac image reconstruction. This approach could provide complementary information that enhances the overall reconstruction process and leads to more comprehensive and detailed images. Furthermore, the development of interpretable deep learning models for cardiac imaging reconstruction could be a valuable direction for future research. By designing models that provide insights into the decision-making process, researchers can enhance the transparency and trustworthiness of reconstruction methods. Interpretable models can help clinicians better understand the reasoning behind the reconstructed images, leading to improved diagnostic accuracy and clinical decision-making.

## CONCLUSION

The study's main findings demonstrate the superior performance of the proposed method in cardiac cine MRI reconstruction, showing improvements in SSIM and PSNR compared to existing models, particularly at accelerated rates. The method's ability to produce high-quality images with fewer artifacts and richer details highlights its potential impact on clinical applications, where accurate and clear imaging is crucial for diagnosis and treatment planning. Furthermore, the emphasis on utilizing spatial-temporal information in dynamic sequences and the effectiveness of the MGDA and MRF modules underscore the contribution of this research to advancing MRI reconstruction techniques. The study's results suggest promising implications for future research directions in improving MRI reconstruction methods for enhanced clinical outcomes and patient care.

# ACKNOWLEDGMENTS

The authors would like to thank the anonymous reviewers and the associate editor for their constructive comments and suggestions that helped to improve both the technical content and the presentation quality of this paper. This work is supported by the National Natural Science Foundation of China under grant No. 61801288.

## Declarations

### Competing interests

The authors declare that they have no known competing financial interests or personal relationships that could have appeared to influence the work reported in this paper. The authors declare the following financial interests/personal relationships which may be considered as potential competing interests: Qiaohong Liu reports financial support was provided by National Natural Science Foundation of China.

### Authors' contributions

Xiaoxiang Han wrote the main part of the paper and conducted most of the experiments. Qiaohong Liu is the project leader who supervised the experiments and paper writing. Yiman Liu and Yang Chen organized and analyzed all the data. Keyan Chen and Yuanjie Lin conducted part of the experiments. Weikun Zhang created all the figures. All authors reviewed the manuscript.

**Funding**

This work is supported by the National Natural Science Foundation of China under grant No. 61801288.

**Availability of data and materials**